\documentclass[10pt,a4paper]{article}

\usepackage[utf8]{inputenc} 
\usepackage[english]{babel} 

\usepackage{graphicx}
\usepackage{mathtools, cuted}
\usepackage{hyperref}

\usepackage{mathtools}
\usepackage{amsmath, amssymb}

\usepackage{color}

\newtheorem{theorem}{Theorem}[section]
\newtheorem{definition}[theorem]{Definition}

\newtheorem{lemma}[theorem]{Lemma} 
 
\newtheorem{remark}[theorem]{Remark}

\newcommand{\qed}{\nobreak \ifvmode \relax \else \ifdim\lastskip<1.5em \hskip-\lastskip \hskip1.5em plus0em minus0.5em \fi \nobreak \vrule height0.75em width0.5em depth0.25em\fi}

\usepackage{geometry}
\makeatletter
\newcommand*\titleheader[1]{\gdef\@titleheader{#1}}
\AtBeginDocument{%
  \let\st@red@title\@title
  \def\@title{%
    \bgroup\normalfont\large\centering\@titleheader\par\egroup
    \vskip1.5em\st@red@title}
}
\makeatother

\title{\LARGE \bf Mathematical Modeling of Guidance Trajectory with a Moving Destination Using Conditionally Markov Modeling}
\titleheader{}
\author{\large{Reza Rezaie and X. Rong Li}
\date{}
\thanks{The authors are with the Department of Electrical Engineering, University of New Orleans, New Orleans, LA 70148. Email addresses are
 {\tt\small rrezaie@uno.edu} and {\tt\small xli@uno.edu}. }
 \thanks{Research supported by NASA through grant NNX13AD29A.}
}

\begin{document}

\maketitle

\thispagestyle{plain}
\pagestyle{plain}

\begin{abstract}

A trajectory of a destination-directed moving object (e.g. an aircraft from an origin airport to a destination airport) has three main components: an origin, a destination, and motion in between. We call such a trajectory that end up at the destination \textit{destination-directed trajectory (DDT)}. A class of conditionally Markov (CM) sequences (called CM$_\text{L}$) has the following main components: a joint density of two endpoints and a Markov-like evolution law. A CM$_\text{L}$ dynamic model can describe the evolution of a DDT but not of a guided object chasing a moving guide. The trajectory of a guided object is called a \textit{guided trajectory (GT)}. Inspired by a CM$_\text{L}$ model, this paper proposes a model for a GT with a moving guide. The proposed model reduces to a CM$_\text{L}$ model if the guide is not moving. We also study filtering and trajectory prediction based on the proposed model. Simulation results are presented.

\end{abstract}


\section{Introduction}

Consider the problem of trajectory modeling with destination information, e.g., a flight from an origin to a destination. Such a problem has three main components: an origin, a destination, and motion in between. A Markov process, which can be described by an evolution law and an initial density, does not fit DDTs because it does not take the destination information into account in general. The CM$_\text{L}$ sequence \cite{CM_I_Conf} has three main components: a joint endpoint density (in other words, an origin density and a destination density conditioned on the origin) and a Markov-like evolution law. CM$_\text{L}$ sequences naturally fits the main DDT components \cite{DD_Conf}. However, they cannot model GT with a moving guide.

Trajectory modeling and trajectory prediction with an intent or a destination have been studied in the literature. Some papers use a combination of the existing trajectory models without a destination and some heuristic modifications to handle trajectory modeling and prediction with destination information. These papers use estimation approaches developed for the case of no intent or destination and utilize intent/destination information to improve trajectory prediction \cite{Hwang0}--\cite{Krozel2}. In \cite{Hwang0}--\cite{Louis} the authors presented some trajectory prediction approaches based on hybrid estimation aided by intent information for air traffic control (ATC). In \cite{Destination_Constraint_1}, the authors used a pseudo measurement approach to incorporate destination information and improve estimation results. In \cite{Krozel}--\cite{Krozel2}, a correlation factor and ADSB intent information was utilized to improve trajectory filtering and prediction in ATC. The above approaches are mainly based on a Markov model aided by some heuristic modifications due to the destination. But such models are not mathematically solid and hard to analyze. To study, generate, and analyze trajectories, it is desired to have a good mathematical model of trajectories providing a solid foundation for further studies. 

In some papers on trajectory modeling and prediction, trajectories are explicitly modeled without using heuristics. Due to uncertainty, trajectories are mathematically modeled by stochastic processes. Consider stochastic sequences defined over the time interval $[0,N]=( 0,1,\ldots,N )$. A sequence is Markov if and only if (iff) conditioned on the state at any time $j$, the segment before $j$ is independent of the segment after $j$. A sequence is reciprocal iff conditioned on the states at any two times $j$ and $l$, the segment inside the interval $(j,l)$ is independent of the segments outside $[j,l]$. In other words, inside and outside $[j,l]$ are independent given the boundaries. A sequence is CM$_\text{L}$ iff conditioned on the state at time $N$, the sequence is Markov over $[0,N-1]$ \cite{CM_I_Conf}. Every Markov sequence is a reciprocal sequence (RS) and every RS is a CM$_\text{L}$ sequence. 

In \cite{Ship} the problem of incorporating predictive information in a Markov model was considered. In \cite{S_1}--\cite{S_2}, the authors presented an approach for intent inference based on bridging distributions. That approach can be seen in a reciprocal process (RP) setting, although RPs were not explicitly used or mentioned in \cite{S_1}--\cite{S_2}. Considering quantized state space, \cite{Fanas1}--\cite{Fanas2} used finite-state RSs in intent inference and tracking. RPs are interesting for modeling trajectories with a destination. But it is not always feasible or easy to quantize the state space. So, it is desired to model trajectories as continuous-state, such as Gaussian sequences. In \cite{Levy_Dynamic}, a dynamic model of nonsingular Gaussian (NG) RSs was presented. However, there are some difficulties about that model and its extensions \cite{White_Gaussian}. For example, due to the nearest-neighbor structure and dynamic noise correlation, state estimation based on that model is not straightforward and several papers \cite{Levy_Smooth}--\cite{Moura1} were devoted to it. In \cite{DD_Conf}, we presented a model for DDTs using a CM$_\text{L}$ model with white dynamic noise.

GT modeling is important in many applications, including biology, robotics, aerospace, pursuit and evasion, traffic control, and autonomous vehicles. Pursuit and evasion behavior, widely observed in nature, has a very important role in predator foraging, prey survival, mating, and territorial battles in species \cite{Leonard}. In robotics, pursuit and evasion games were used to study motion planning problems \cite{Search_robot}. Trajectory of a vehicle pursuing another one on a street is also a GT with a moving guide. GTs with a moving guide can also be found in a problem with a team of unmanned aerial and ground vehicles pursuing another team of evaders \cite{Team_UAV}. Modeling GTs with a moving guide is important in guidance, homing, and interception in aerospace applications \cite{Aero_DDT_1}--\cite{Aero_DDT_3}. An example is a moving object intercepting another moving object. A systematic approach for modeling a GT with a moving guide in an appropriate mathematical model is desired.

Gaussian CM processes were introduced in \cite{Mehr}. Inspired by \cite{Mehr}, we presented definitions of different classes of (Gaussian/non-Gaussian) CM processes (including CM$_\text{L}$) in \cite{CM_I_Conf}, where NG CM sequences were studied, modeled and characterized \cite{Thesis_Reza}. Also, a dynamic model with white dynamic noise, called a CM$_\text{L}$ model, was presented for state evolution of NG CM$_\text{L}$ sequences. As a special case of the CM$_\text{L}$ model, a reciprocal CM$_\text{L}$ model with white noise was presented in \cite{CM_II_A_Conf}. In \cite{CM_III_Journal}, we presented the notion of a Markov-induced CM$_\text{L}$ model as a tool for application of CM$_\text{L}$ sequences.

The main contributions of this paper are as follows. We propose a model to describe a GT with a moving guide. The model is a generalization of our CM$_\text{L}$ model for DDTs presented in \cite{DD_Conf}. We discuss parameter design of the proposed model. Also, we derive trajectory filters and predictors based on the proposed trajectory model.

The paper is organized as follows. In Section \ref{DD_Modeling}, a DDT modeling using CM$_\text{L}$ sequences is reviewed. In Section \ref{DD_Moving}, our proposed model for a GT with a moving guide is presented. Section \ref{Filtering} presents the corresponding trajectory filter and predictor. Simulation examples are presented in Section \ref{Simulation}. Section \ref{Conclusion} contains conclusions.

\section{DDT Modeling Using CM$_\text{L}$ Sequences}\label{DD_Modeling}

The following notation is used:
\begin{align*}
[i,j]& \triangleq ( i,i+1,\ldots ,j-1,j ), \quad i<j\\
[x_k]_{i}^{j} & \triangleq ( x_i, x_{i+1}, \ldots, x_j )\\
[x_k] &  \triangleq [x_k]_{0}^{N}
\end{align*}
where $k$ is the discrete-time index. $C_{i,j}$ is a covariance function and $C_k \triangleq C_{k,k}$. Also, $F(\cdot | \cdot)$ is the conditional cumulative distribution function (CDF) and $p(\cdot | \cdot)$ is a conditional density. The symbol ``$ ' $" stands for matrix transposition. ZMNG and NG stand for ``zero-mean nonsingular Gaussian" and ``nonsingular Gaussian", respectively. $\mathbb{R}$ denotes the set of real numbers. $\mathcal{N}(\mu , C)$ is the Gaussian distribution with mean $\mu$ and covariance $C$. $\mathcal{N}(x;\mu , C)$ is the corresponding Gaussian density.

\begin{definition}\label{Markov_CDF}
$[x_k]$ is Markov if\footnote{$F(\xi_k|x_j)=P\lbrace x^1_k\leq \xi^1_k, x^2_k\leq \xi^2_k, \ldots , x^d_k\leq \xi^d_k|x_j \rbrace$, where for example $x^1_k$ and $\xi^1_k$ are the first entries of the vectors $x_k$ and $\xi_k$, respectively. Likewise for other CDFs.} 
 \begin{align}
 F(\xi _k|[x_{i}]_{0}^{j})=F(\xi _k|x_j)
 \end{align} 
$\forall j<k$, $\forall \xi _k \in \mathbb{R}^d$, where $d$ is the dimension of $x_k$.  

\end{definition}

\begin{lemma}\label{Model_Dynamic_Proposition}
A ZMNG $[x_k]$ is Markov iff
\begin{align}
x_k=M_{k,k-1}x_{k-1}+w_{k}, k \in [1,N], \quad x_0=w_0 \label{Markov_Model}
\end{align}
where $[w_k]$ ($M_k=\text{Cov}(w_k)$) is a ZMNG white sequence.

\end{lemma}

Without the notion of destination the trajectory of a moving object has two main elements: an origin and an evolution law. A Markov sequence is determined by two elements: an initial density and an evolution law. Sample paths of a Markov sequence can model such trajectories. A Markov sequence, with its final density uniquely determined by its initial density and its evolution law, is not powerful or flexible enough for DDT modeling.

\subsection{CM$_\text{L}$ Sequences for DDT Modeling}\label{CML_DD}

\begin{definition}[\cite{CM_I_Conf}]\label{CMc_CDF}
$[x_k]$ is CM$_\text{L}$ if
 \begin{align}
 F(\xi _k|[x_{i}]_{0}^{j},x_{N})=F(\xi _k|x_j,x_N)\label{CDF_1}
 \end{align} 
$\forall j,k \in [0,N]$, $j<k$, $\forall \xi _k \in \mathbb{R}^d$.

\end{definition}

A CM$_\text{L}$ model of the ZMNG CM$_\text{L}$ sequence is as follows.
\begin{theorem}[\cite{CM_I_Conf}]\label{CML_Forward_Dynamic_Proposition}
A ZMNG $[x_k]$ is CM$_\text{L}$ iff it obeys
\begin{align}
x_k=G_{k,k-1}x_{k-1}+G_{k,N}x_N+e_k, \quad k \in [1,N-1] 
\label{CML_Dynamic_Forward}
\end{align}
where $[e_k]$ ($\text{Cov}(e_k)=G_k$) is a ZMNG white sequence, and boundary condition
\begin{align}
&x_0=e_0, \quad x_N=G_{N,0}x_0+e_N \label{CML_Forward_BC1}
\end{align}

\end{theorem}

Reciprocal sequences are special CM$_\text{L}$ sequences.

\begin{lemma}[\cite{CM_II_A_Conf}]\label{CDF}
$[x_k]$ is reciprocal iff 
\begin{align}
F(\xi _k|[x_{i}]_{0}^{j},[x_i]_l^N)=F(\xi _k|x_j,x_l)
\end{align}
$\forall j,k,l \in [0,N]$ ($j < k < l$), $\forall \xi _k \in \mathbb{R}^d$.  

\end{lemma}

Some desirable properties of CM$_\text{L}$ sequences for DDT modeling are: 1) they model the main DDT components well, 2) they have a Markov-like evolution law, which is simple and well understood, 3) the CM$_\text{L}$ model can systematically model the impact of destination on the evolution of trajectories, 4) the CM$_\text{L}$ model has desirable white dynamic noise, 5) state estimation based on the CM$_\text{L}$ model is straightforward, and 6) CM$_\text{L}$ sequences (and their models) can be easily and systematically generalized, if necessary. However, the CM$_\text{L}$ modeling of a DDT assumes a fixed destination and cannot model trajectories with a moving destination.

The structure of our CM$_\text{L}$ model and that of the reciprocal model of \cite{Levy_Dynamic} for DDT modeling are compared as follows. The dynamic model presented in \cite{Levy_Dynamic} has a nearest-neighbor structure, that is, the current state depends on the previous state and the next state. So, information (density) of the next state is needed for estimation of the current state, but such information is not available. Based on our CM$_\text{L}$ model, information (density) about the last state (destination) is required and available for DDT modeling in practice. In addition, the dynamic noise in the model of \cite{Levy_Dynamic} is colored, which makes state estimation not straightforward. By contrast, the dynamic noise of our CM$_\text{L}$ model is white and its state estimation is straightforward.

For trajectory modeling we need nonzero-mean sequences. A nonzero-mean NG sequence is CM$_\text{L}$ (or Markov) iff its zero-mean part follows a CM$_\text{L}$ model of Theorem \ref{CML_Forward_Dynamic_Proposition} (or Lemma \ref{Model_Dynamic_Proposition}). A CM$_\text{L}$ model of nonzero-mean Gaussian CM$_\text{L}$ sequences for DDT modeling is as follows. Let $\mu _0$ ($\mu _N$) and $C_0$ ($C_N$) be the mean and covariance of the origin (destination) state. Also, let $C_{0,N}$ be the cross-covariance of $x_0$ and $x_N$. We have $\eqref{CML_Dynamic_Forward}$ and
\begin{align}
x_N&=\mu_N + G_{N,0}(x_0-\mu_0) + e_N\label{CML_Non_2}\\
x_0&=\mu _0 + e_0\label{CML_Non_3}
\end{align}
where $e_k \sim \mathcal{N}(0,G_k), k \in [0,N]$, $G_{N,0}=C_{N,0}C_0^{-1}$, $G_N=\text{Cov}(e_N)=C_N-C_{N,0}C_0^{-1} (C_{N,0})'$, and $G_0=C_0$.

\subsection{Parameter Design of CM$_\text{L}$ Model for DDT}\label{CML_DD_Parameters}

A DDT can be modeled based on two key assumptions \cite{CM_III_Journal}: (a) the motion follows a Markov model $\eqref{Markov_Model}$ (e.g., a nearly constant velocity model) without considering the destination information, and (b) the joint origin and destination density is known (exactly or approximately). Now, let $[s_k]$ be Markov modeled by $\eqref{Markov_Model}$. Since every Markov sequence is CM$_\text{L}$, $[s_k]$ can be modeled by a CM$_\text{L}$ model $\eqref{CML_Dynamic_Forward}$ with ($G^s_N=\text{Cov}(e_N)$, $G^s_0=\text{Cov}(e_0)$)
 \begin{align}
s_0&=e_0, \quad s_N=G^{s}_{N,0}s_0+e_N \label{CML_R_FQ_BC2}
\end{align}
where by $p(s_k|s_{k-1},s_N)=\frac{p(s_k|s_{k-1})p(s_N|s_k,s_{k-1})}{p(s_N|s_{k-1})}=\mathcal{N}(s_k;G_{k,k-1}s_{k-1}+G_{k,N}s_N,G_k)$ and the Markov property, $\forall k \in [1,N-1]$, we have
\begin{align}
G_{k,k-1}&=M_{k,k-1}-G_{k,N}M_{N|k}M_{k,k-1} \label{CML_Choice_1}\\
G_{k,N}&=G_kM_{N|k}'C_{N|k}^{-1} \label{CML_Choice_2}\\
G_k&=(M_k^{-1}+M_{N|k}'C_{N|k}^{-1}M_{N|k})^{-1}\label{CML_Choice_3}
\end{align}
where\footnote{By matrix inversion lemma, $\eqref{CML_Choice_3}$ is equivalent to $G_k=M_{k} - M_{k}M_{N|k}'(C_{N|k} + M_{N|k}M_{k}M_{N|k}')^{-1}M_{N|k}M_{k}$.} $M_{N|N}=I$,
\begin{align*}
M_{N|k}&=M_{N,N-1}\cdots M_{k+1,k}, \quad k \in [1,N-1]\\
C_{N|k}&=\sum _{n=k}^{N-1} M_{N|n+1}M_{n+1} M_{N|n+1}', \quad k \in [1,N-1]
\end{align*}
and $M_{k,k-1}, M_k, k \in [1,N]$, are parameters of $\eqref{Markov_Model}$.

Now, consider a different sequence $[x_k]$ described by the same evolution model $\eqref{CML_Dynamic_Forward}$ but a different boundary condition $\eqref{CML_Forward_BC1}$ with ($\text{Cov}(e_N)$, $G_{N,0}$, $\text{Cov}(e_0))=$ $(G_N,G_{N,0},G_0) \neq (G^s_N,G^s_{N,0},G^s_0)$. So, $[s_k]$ and $[x_k]$ are two different sequences. By Theorem \ref{CML_Forward_Dynamic_Proposition}, $[x_k]$ is a CM$_\text{L}$ sequence. The sequences $[s_k]$ and $[x_k]$ have the same CM$_\text{L}$ evolution model/law (i.e., have the same parameters $G_{k,k-1}, G_{k,N},$ $ G_k, k \in [1,N-1]$), but $[x_k]$ can have any joint endpoint density since parameters of its boundary condition (i.e., $(G_N,G_{N,0},G_0)$) are arbitrary.

The CM$_\text{L}$ model $\eqref{CML_Dynamic_Forward}$ with parameters $\eqref{CML_Choice_1}$--$\eqref{CML_Choice_3}$ is called the CM$_\text{L}$ model \textit{induced} by the Markov model $\eqref{Markov_Model}$ (or simply the \textit{Markov-induced} CM$_\text{L}$ model) since its parameters are obtained from parameters of $\eqref{Markov_Model}$.

\section{Modeling GT with a Moving Guide}\label{DD_Moving}


A DDT can be naturally modeled as a CM$_\text{L}$ sequence $[x_k]$ as follows. The origin, the destination, and their relationship are modeled by a joint density of $x_0$ and $x_N$. For a DDT, the density of $x_N$ is (assumed) known. So, the evolution law is modeled as a density conditioned on $x_N$. The simplest such density is the product of its marginals: $p([x_k]_0^{N-1}|x_N)=\prod _{k=0}^{N-1}p(x_k|x_N)$. But this conditionally independent law is often inadequate. Then, the next choice is often a CM density: $p([x_k]_0^{N-1}|x_N)=p(x_0|x_N)\prod _{k=1}^{N-1}p(x_k|x_{k-1},x_N)$, which is the evolution law of a CM$_\text{L}$ sequence. The main components of a CM$_\text{L}$ sequence $[x_k]$ are: a CM evolution law (conditioned on $x_N$) and a joint density of $x_0$ and $x_N$. Similarly, we can consider more general and complicated evolution laws (e.g., higher-order CM densities), if needed or desired. Therefore, by choosing conditional laws, a DDT can be modeled well. 

An illustrative application of GT with a moving guide is guidance, where object A (e.g., a missile) is chasing (i.e., guided by) a moving object B (e.g., an aircraft). Object A is a guided object and object B is a moving guide.

The above argument for DDTs does not work straightforwardly for a GT with a moving guide. It is explained as follows. For a GT with a moving guide, the origin can be modeled similar to that of a DDT. However, since the guide is moving (time varying), a random variable can not model it. Therefore, the motion of a \textit{moving guide} is modeled by a stochastic sequence. So, the whole problem, including the trajectory of the guided object and its moving guide, can be modeled as follows. A GT is modeled by a sequence $[x_k]$ and its moving guide trajectory is modeled by a sequence $[d_k]$. Then, relationship between the two sequences should be determined. Following this idea, a model is presented for a GT with a moving guide below.

Assume there is a moving object chasing a moving guide. The goal is to model the trajectory of the object (i.e., GT) and its moving guide. Let the state evolution of the moving guide be modeled by a Markov sequence $[d_k]$. So, we have
\begin{align}
d_k=G^d_{k,k-1}d_{k-1}+w_{k}, k \in [1,N], \quad d_0=w_0\label{Evador}
\end{align}
where $[w_k]$ is a white Gaussian sequence.

$\eqref{Evador}$ can be any motion model (e.g., nearly constant velocity/acceleration/turn model).

A GT is modeled by a sequence $[x_k]$ described by
\begin{align}
x_k&=G^x_{k,k-1}x_{k-1}+G^{xd}_{k,k-1}d_{k-1}+e_{k}, k \in [1,N-1]\label{Pursuer}\\
x_0&=e_0\\
x_N&=d_N+e_N
\end{align}
where $[e_k]$ is a white Gaussian sequence uncorrelated with $[w_k]$.

The above model for GT is justified as follows. At time $k-1$ the object sets $d_{k-1}$ as its guide. Then, following the idea of DDT, the state evolution of the object at the moment is modeled by a CM$_\text{L}$ model considering the guide as its destination at the moment. The dynamic noise $e_{k}$ is used to model all stochastic deviations from the deterministic relationship (i.e., $\eqref{Pursuer}$ without $e_k$). The state evolution for the next time is justified in the same way.  

The underlying idea behind $\eqref{Pursuer}$ is informally explained as follows. A random variable can model a time-invariant phenomenon/effect. To describe a time-varying phenomenon/effect, a stochastic process is needed. To model a GT with a moving guide, the final state of a CM$_\text{L}$ sequence in its dynamic model is replaced by a stochastic sequence that models the state evolution of the moving guide. Since the trajectory of the moving guide is Markov in $\eqref{Evador}$, the GT model $\eqref{Pursuer}$ is called a \textit{Markov-guided} CM$_\text{L}$-like model. It reduces to the model $\eqref{CML_Dynamic_Forward}$ if the guiding sequence $[d_k]$ reduces to the final state $x_N$ of the guided sequence (i.e., $[d_k]=x_N$).

It is well known how to determine parameters of the Markov model $\eqref{Evador}$ of the moving guide, e.g., as a nearly constant velocity/acceleration/turn model \cite{T_2}. To determine parameters of $\eqref{Pursuer}$, we follow the idea of a Markov-induced CM$_\text{L}$ model for a DDT: At any time we assume a moving guide to be fixed and treat the corresponding dynamic model $\eqref{Pursuer}$ as a DDT model. So, like the DDT model, parameters of $\eqref{Pursuer}$ are determined for that time based on the idea of a Markov-induced CM$_\text{L}$ model. For simplicity, consider a time-invariant Markov model $\eqref{Markov_Model}$ (e.g., a nearly constant velocity) with $M_{k,k-1}=F$ and $M_k=Q$. Then, by $\eqref{CML_Choice_1}$--$\eqref{CML_Choice_3}$, the parameters of the model $\eqref{Pursuer}$ are determined as 
\begin{align}
G_{k,k-1}&=F-G_{k,N}F^{N-k+1} \label{CML_Choice_1_F}\\
G_{k,N}&=G_k(F^{N-k})'C_{N|k}^{-1} \label{CML_Choice_2_F}\\
G_k&=(Q^{-1}+(F^{N-k})'C_{N|k}^{-1}F^{N-k})^{-1}\label{CML_Choice_3_F}
\end{align}
where $C_{N|k}=\sum _{i=0}^{N-k-1}F^iQ(F^i)'$, $k \in [1,N-1]$.

\begin{remark}
$\eqref{Pursuer}$ models a GT $[x_k]$, where there is no destination for the moving guide (or information about it). It is possible to extend model $\eqref{Pursuer}$ to the case where the moving guide has its own destination, i.e., the trajectory of the moving guide is destination-directed, resulting in a CM$_\text{L}$-guided CM$_\text{L}$-like model. Assume that the moving guide sequence $[d_k]$ has its own destination $d_N$. Then, the trajectory of the object and its moving guide are modeled as
\begin{align}
x_k &=G^{x}_{k,k-1}x_{k-1} + G^{xd}_{k,k-1}d_{k-1}+e_{k}, k \in [1,N-1]\label{x}\\
x_0&=e_0\\
x_N&=d_N+e_N\\
d_k&=G^d_{k,k-1}d_{k-1} + G^{dd}_{k,N}d_N + e^d_{k}, k \in [1,N-1]\label{y}\\
d_N&=G^{dd}_{N,0}d_0+e^d_N\label{BC1_y}\\
d_0&=e^d_0
\end{align}
where $[e_k]$ and $[e^d_k]$ are white Gaussian sequences, uncorrelated with each other. 

In the above models, for simplicity, we assumed that the object reaches the guide at time $N$. However, it is also possible to model trajectories where the object reaches the guide sooner than time $N$. 

\end{remark}

\section{Trajectory Filtering and Prediction}\label{Filtering} 

Assume an object going towards its moving guide, where a sensor makes measurements of both the object and its guide. The goal is to estimate the state of both the object and its guide. Since the object and its guide are related, measurements for each one is useful for estimating the states of both. So, both states should be estimated together using measurements of the object and its guide.   

The measurements of the object and its guide are
\begin{align}
z^x_k&=H_kx_k+v^x_k\label{Measurements_x}\\
z^d_k&=H_kd_k+v^d_k\label{Measurements_y}
\end{align}
where the measurement noise $[v^x_k]_1^N$ and $[v^d_k]_1^N$ are zero-mean white, uncorrelated, and uncorrelated with $[w_k]$ and $[e_k]$ of $\eqref{Evador}$--$\eqref{Pursuer}$.

Combining models $\eqref{Evador}$ and $\eqref{Pursuer}$, we have the Markov model 
\begin{align}
s_k&= G_{k,k-1}s_{k-1} + e^s_k\label{Joint_Model}\\
z_k&=H_ks_k+v_k\label{Measurements}
\end{align}
where
\begin{align*}
s_k&=\left[ \begin{array}{c}
x_k\\
d_k
\end{array}\right], e^s_k= \left[ \begin{array}{c}
e_{k}\\
w_{k}
\end{array}\right], \text{Cov}(e^s_k)=G^s_k\\
G^s_{k,k-1}&=\left[ \begin{array}{cc}
G^{x}_{k,k-1} & G^{xd}_{k,k-1}\\
0 & G^{d}_{k,k-1}
\end{array} \right]\\
H_k&=\left[ \begin{array}{cc}
H^x_k & 0\\
0 & H^d_k
\end{array}\right], z_k=\left[ \begin{array}{c}
z^x_k\\
z^d_k
\end{array}\right], v_k= \left[ \begin{array}{c}
v^x_k\\
v^d_k
\end{array}\right] 
\end{align*}

The goal is to obtain $\hat{x}_k=E[x_k|z^k]$ and $\hat{d}_k=E[d_k|z^k]$ and their mean square error (MSE) matrix given all the measurements from the beginning to time $k$ denoted as $z^{k}=\lbrace z_1, z_2, \ldots, z_k\rbrace$. $[e^s_k]$ and $[v_k]_1^N$ are white Gaussian sequences, uncorrelated with each other. By $\eqref{Joint_Model}$--$\eqref{Measurements}$, $s_k$ and $z_k$ are linear combinations of $[e^s_k]$ and $[v_k]_1^N$. So, $s_k$ and $z_k$ are jointly Gaussian, and the minimum mean square error (MMSE) estimate of the states of the object and its guide are
\begin{align}
\hat{s}_k&=E[s_k|z^k]=\hat{s}_{k|k-1}+C_{s_k,z_k}C_{z_k}^{-1}\Big(z_k-H_k\hat{s}_{k|k-1}\Big)\label{s_h}\\
\Sigma _k&=E[(s_k-\hat{s}_k)(s_k-\hat{s}_k)']=\Sigma_{k|k-1}-C_{s_k,z_k}C_{z_k}^{-1}(C_{s_k,z_k})'\label{P_s_h}
\end{align}
where 
\begin{align*}
\hat{s}_{k|k-1}&= G^s_{k,k-1}\hat{s}_{k-1}\\
\Sigma_{k|k-1}&= G^s_{k,k-1}\Sigma _{k-1}(G^s_{k,k-1})'+G^s_{k}\\
C_{s_k,z_k}&=\Sigma _{k|k-1}(H_k)'\\
C_{z_k}&=H_k\Sigma _{k|k-1}(H_k)'+R_k
\end{align*}
and the estimate of $x_k$ and its MSE matrix are
\begin{align*}
\hat{x}_k&=[I ,0]\hat{s}_k\\
P^x_k&=[I ,0]\Sigma _k [I , 0]'
\end{align*}
and the estimate of $d_k$ and its MSE matrix are
\begin{align*}
\hat{d}_k&=[0,I]\hat{s}_k\\
P^d_k&=[0,I]\Sigma _k [0,I]'
\end{align*}

Trajectory is predicted as follows. Let $[s_k]$ be modeled by $\eqref{Joint_Model}$. Assume that the output of the filter $p(s_k|z^k)=\mathcal{N}(s_k;\hat{s}_k,\Sigma_k)$ at time $k$ is available. For $k+n \in [k+1,N-1]$, the prediction and its MSE matrix are obtained as 
\begin{align}
&\hat{s}_{k+n|k}=G^s_{k+n|k}\hat{s}_{k}\label{s_k+n}\\
&\Sigma _{k+n|k}=C_{k+n|k}+G^s_{k+n|k}\Sigma _{k}(G^s_{k+n|k})'\label{S_k+n}
\end{align}
where $G^s_{k|k}=I, \forall k$, and
\begin{align*}
G^s_{k+n|k}&=G^s_{k+n,k+n-1}G^s_{k+n-1,k+n-2} \cdots G^s_{k+1,k}\\
C_{k+n|k}&=\sum _{i=k}^{k+n-1}G^s_{k+n|i+1}G^s_i(G^s_{k+n|i+1})'
\end{align*}
Then, the prediction of $x_{k+n}$ and its MSE matrix are
\begin{align}
&\hat{x}_{k+n|k}=[I , 0]\hat{s}_{k+n|k}\label{x_k+n_2}\\
& P^x_{k+n|k}=[I , 0]\Sigma _{k+n|k}[I , 0]'\label{P^x_k+n_2}
\end{align}
The prediction of $d_{k+n}$ and its MSE matrix are
\begin{align}
&\hat{d}_{k+n|k}=[0 , I]\hat{s}_{k+n|k}\label{y_k+n_2}\\
& P^d_{k+n|k}=[0 , I]\Sigma _{k+n|k}[0 , I]'\label{P^y_k+n_2}
\end{align}

Filtering and prediction based on $\eqref{x}$--$\eqref{y}$ are similar. Combining $\eqref{x}$ and $\eqref{y}$, we have the Markov model 
\begin{align}
s_k&= G^s_{k,k-1}s_{k-1} + e^s_k\label{Joint_Model_2}\\
z_k&=H_ks_k+v_k\label{Measurements_2}
\end{align}
where 
\begin{align*}
G^s_{k,k-1}&=\left[ \begin{array}{ccc}
G^{x}_{k,k-1} & G^{xd}_{k,k-1} & 0\\
0 & G^{d}_{k,k-1} & G^{dd}_{k,N}\\
0 & 0 & I
\end{array} \right]\\
s_k&=\left[ \begin{array}{c}
x_k\\
d_k\\
d_N
\end{array}\right], e^s_k= \left[ \begin{array}{c}
e^x_{k}\\
e^d_{k}\\
0
\end{array}\right]\\ 
H_k&=\left[ \begin{array}{ccc}
H^x_k & 0 & 0\\
0 & H^d_k & 0
\end{array}\right], z_k=\left[ \begin{array}{c}
z^x_k\\
z^d_k
\end{array}\right], v_k= \left[ \begin{array}{c}
v^x_k\\
v^d_k
\end{array}\right]
\end{align*}
Then, filtering and prediction are based on $\eqref{Joint_Model_2}$--$\eqref{Measurements_2}$.

\section{Simulations}\label{Simulation}

A simulation study of the proposed model for modeling GT with a moving guide is reported in this section. 

Assume a two-dimensional scenario, where the state of a moving object at time $k$ is $x_k=[ \mathsf{x} , \mathsf{\dot{x}}, \mathsf{y} , \mathsf{\dot{y}}]_k'$ with position $[\mathsf{x},\mathsf{y}]'$ and velocity $[\mathsf{\dot{x}},\mathsf{\dot{y}}]'$. The state of the moving guide $d_k$ is defined similarly.

Let the GT of a moving object be modeled by $\eqref{Pursuer}$ and the trajectory of its moving guide by $\eqref{Evador}$. Parameters of the Markov model $\eqref{Evador}$ are $G^d_{k,k-1}=\text{diag}(F_1,F_1)$ and $\text{Cov}(w_{k})=\text{diag}(Q_1,Q_1)$, $k \in [1,N]$, where $F_1=\left [\begin{array}{cc}
1 & T\\
0 & 1
\end{array}\right]$, $Q_1=q\left[
\begin{array}{cc}
T^3/3 & T^2/2\\
T^2/2 & T
\end{array} \right]$, $T=1$ second, $q=0.005$, and $N=250$. Also, parameters of $\eqref{Pursuer}$ are given by $\eqref{CML_Choice_1_F}$--$\eqref{CML_Choice_3_F}$, where $F=\text{diag}(F_1,F_1)$ and $Q=\text{diag}(Q_1,Q_1)$.

\begin{figure}
\centering
    \includegraphics[scale=0.13]{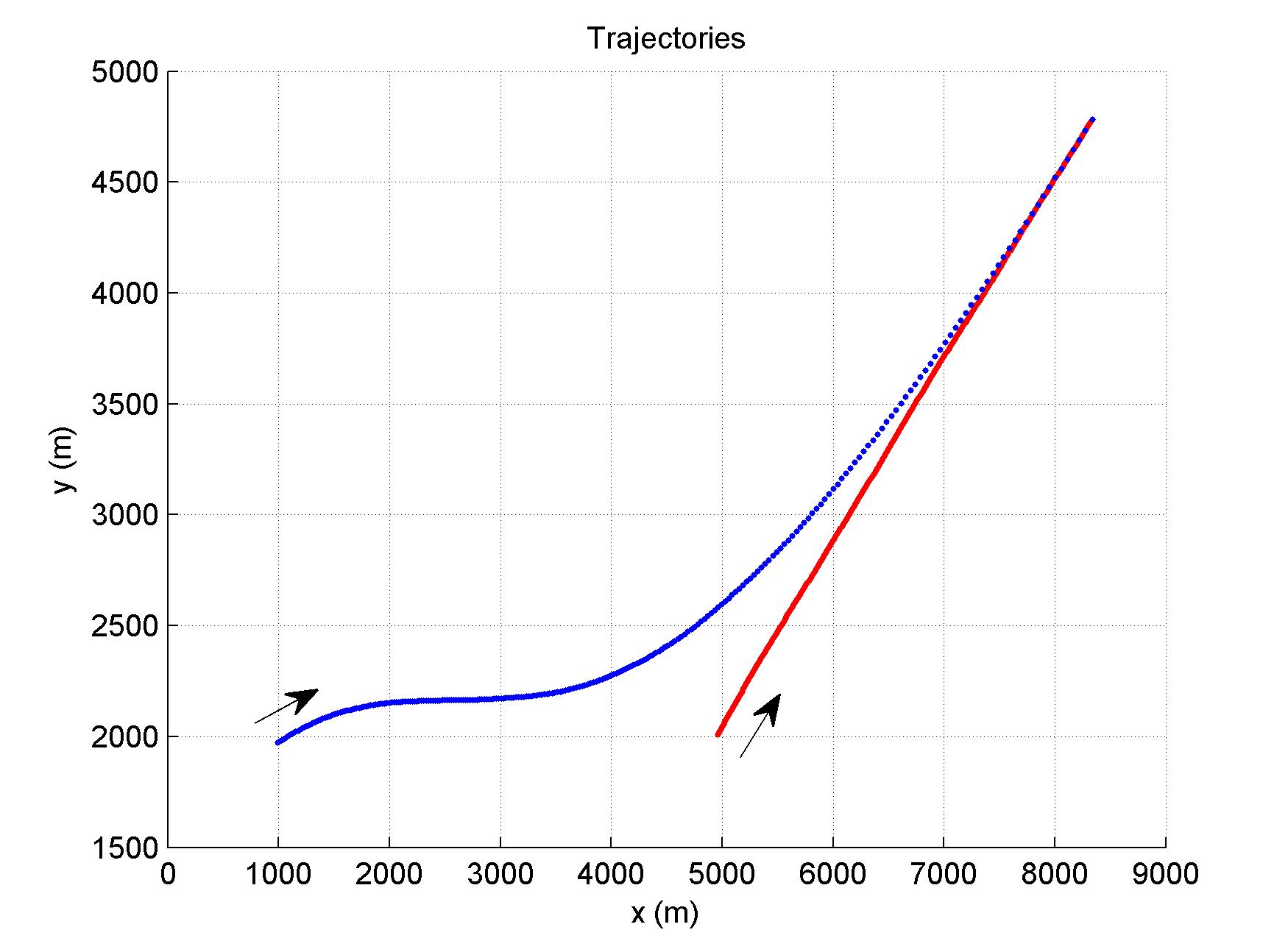}
\caption{Trajectory of a guided object (blue line) chasing its moving guide (red line).}
\label{F1}
\end{figure}

\begin{figure}
\centering
    \includegraphics[scale=0.13]{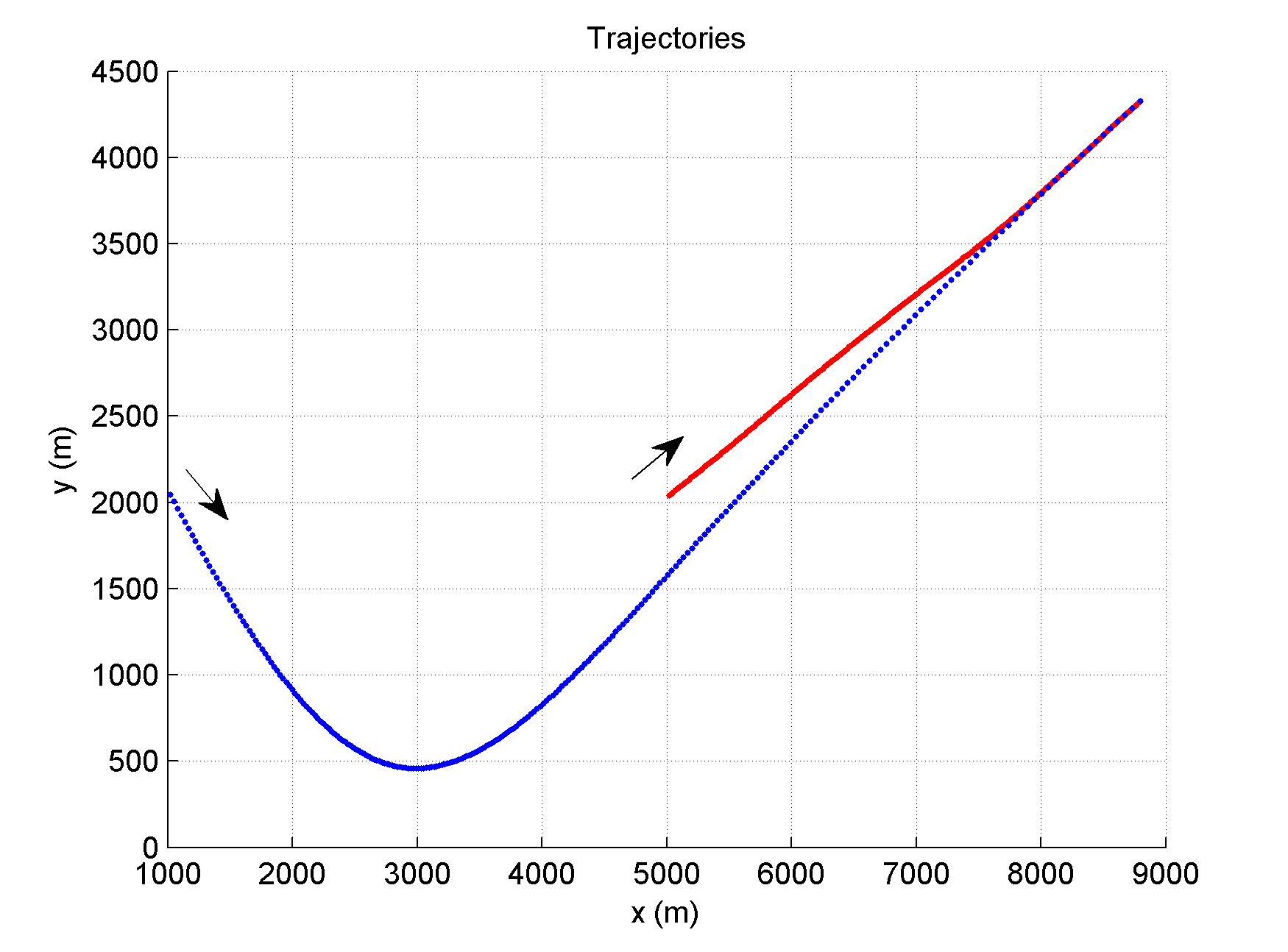}
\caption{Trajectory of a guided object (blue line) chasing its moving guide (red line).}
\label{F2}
\end{figure}

\begin{figure}
\centering
    \includegraphics[scale=0.13]{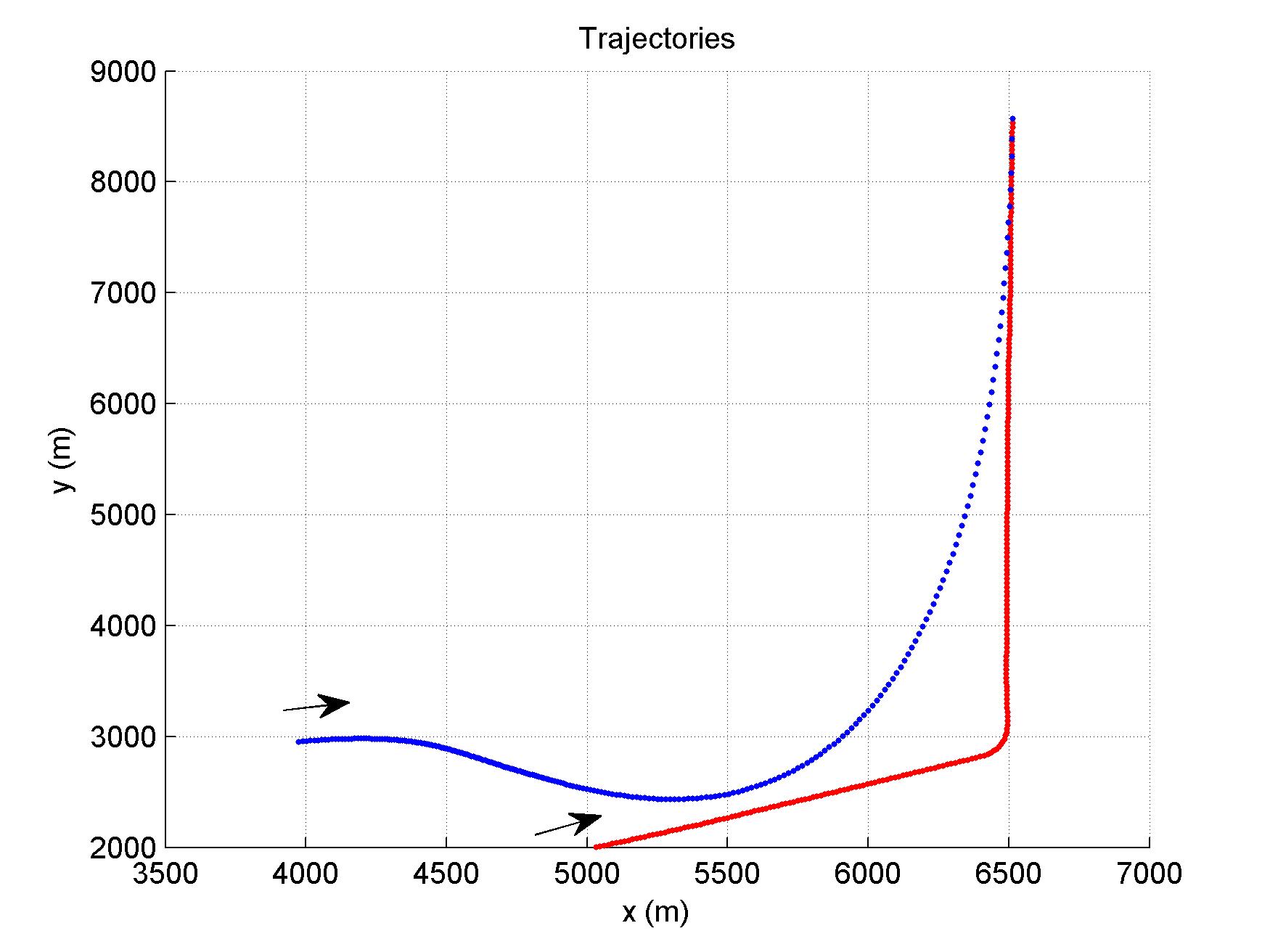}
\caption{Trajectory of a guided object (blue line) chasing its moving guide (red line).}
\label{F3}
\end{figure}

Figs. \ref{F1}, \ref{F2}, \ref{F3} show the GT of an object (blue line) along with that of its moving guide (red line) in three scenarios. They show how the object can chase and reach its moving guide, where there are different initial states of the object and its guide.

\section{Conclusions}\label{Conclusion}

CM$_\text{L}$ sequences model DDTs well, but they cannot model well a GT with a moving guide because they cannot take a moving guide into account. This is because a CM$_\text{L}$ sequence models the destination of a DDT by the final state of the sequence. The final state is a random variable and does not change over time. A stochastic sequence is needed to model the trajectory of a moving guide.

Inspired by a CM$_\text{L}$ dynamic model, a model has been proposed for a GT with a moving guide. Model parameter design and the corresponding optimal trajectory filtering and prediction have been also studied. 

Following the idea of DDT modeling by a CM$_\text{L}$ sequence, the proposed GT model has been extended to the case where the moving guide has its own destination.

CM sequences/models are powerful modeling tools. For example, a CM$_\text{L}$ dynamic model can describe a DDT straightforwardly. Although a CM$_\text{L}$ model cannot directly take a moving guide into account to model a GT, it provides a foundation for a model that can describe GT with a moving guide.


\end{document}